\magnification=\magstep1
\hsize=15.5truecm    
\vsize=22.0truecm
\hoffset=0.5cm
\font\ninerm=cmr9    scaled\magstep0

\font\tit=cmbx10      scaled\magstep1
\font\sect=cmbx9      scaled\magstep1

\nopagenumbers
\pageno=1
\def\today{\sl \ifcase\month\or
January\or February\or March\or April\or May\or June\or
July\or August\or September\or October\or November\or December\fi
\space\number\day, \number\year}
\newcount\nim \nim=0
\def\nitem {\advance\nim by 1 \bigbreak\item{\number\nim .}}
\newcount\nime \nime=0
\def\numeq {\global\advance\nime by 1 \eqno( \number\nime )}
\newcount\nref \nref=0 \newcount\nrefx \nrefx=0
\def\ref#1{{\global\advance\nref by 1}{\global\advance\nrefx by #1}
{\ifnum #1=1{$^{\number\nref)}$}\else{$^{\number\nref-\number\nrefx)}$}\fi}
{\global\nref=\number\nrefx}\hglue-0.2truecm}
\def\b8{~$^{8}$B$\,$}
\def\be7{~$^{7}$Be$\,$}
\def\pb208{~$^{208}$Pb$\,$}

\def\pni{\par\noindent}

\def\etal{{\it et al}}

\def\half{{\scriptstyle{1\over2}}}

\def\makeheadline{\vbox to 0pt{\vskip-40pt
   \line{\vbox to8.5pt{}\the\headline}\vss}\nointerlineskip}
\headline={\ifodd\pageno\rightheadline \else\leftheadline\fi}
\def\rightheadline{\ifnum\pageno=1 {\hfill \hfill}   
 \else {\sl\hfil
 C.H. Dasso, S.M. Lenzi and A. Vitturi, Dominance of nuclear processes...
 \hfil {\bf \folio}}\fi}
\def\leftheadline{\sl\hfil
 C.H. Dasso, S.M. Lenzi and A. Vitturi, Dominance of nuclear processes...
 \hfil {\bf \folio} }
\voffset=4\baselineskip
\baselineskip=13pt
\parskip=0.5truecm
\overfullrule=0pc
\null
\centerline {\sl Version as of \today}
\vskip 1.0truecm
\centerline  {\tit Dominance of Nuclear Processes in the 
              Dissociation of $^{\bf 8}$B}
\vskip 1.2truecm
\centerline {C.H. Dasso $^{\rm a}$, S.M. Lenzi $^{\rm b}$
             and A. Vitturi $^{\rm b}$}
\vskip 0.8truecm
\baselineskip=11pt
\centerline {ECT*, Strada delle Tabarelle 286, {\ninerm I-38050} Trento, Italy}
\centerline {and The Niels Bohr Institute, Blegdamsvej 17,
 Copenhagen \O, Denmark $^{\rm a}$}
\vskip 0.3truecm
\centerline {Dipartimento di Fisica and INFN,
 Universit\`a di Padova, Padova, Italy $^{\rm b}$}
\vskip 1.6truecm
\baselineskip=12pt
\pni
 {\bf Abstract:}
\pni\sl
 We study the break-up of \b8 in collisions with heavy-ions.  
 The process is described in terms of inelastic excitations leading 
 to states in the continuum. 
 The effects of the nuclear and Coulomb fields induced by the reaction 
 partner are included on the same footing in the microscopic construction 
 of the transition matrix elements.\break
 At variance with previous findings, the contribution of the nuclear 
 component is found to be comparable to --~and even larger than~-- that of
 the Coulomb one.
 Because of the weak binding energy of \b8 the nuclear
 couplings associated with inelastic excitation to low-lying states
 in the continuum extend to unusually large distances.
 As a consequence, the interplay between nuclear and Coulomb excitation
 processes differs significantly from the situation encountered in 
 reactions involving systems close to the stability line.  
 In particular, nuclear excitation is found to remain predominant at 
 energies well below the Coulomb barrier.
\vskip 0.3truecm
\pni\rm
\vfill
\eject
\null
\baselineskip=14.1pt
\vskip 2.9truecm
\pni
{\sect 1. Introduction}
\par
 One of the most interesting aspects of nuclear systems at the nucleon 
 drip-lines is the prediction (supported by experimental evidence) 
 of strong concentrations of strength in the multipole response at  
 excitation energies near the continuum threshold for particle emission.
 It has been shown in a series of investigations$^{\,1-3)}$ that this strength 
 is not associated with the existence of a low-lying collective mode
 but, rather, a characteristic feature of the single-particle response 
 for weakly-bound systems.
 This arises from the possibility of setting up a favorable matching 
 between the long range of the bound orbitals and the wavelength of 
 scattering states in the continuum.  
 Although it is not readily appreciated, this argument invokes the
 single-particle transition density and it is not thus restricted to
 electromagnetic processes.
\par
 Conventional wisdom takes for granted that Coulomb excitation has
 a long interaction range while nuclear processes only become important 
 at relative distances comparable to the sum of the nuclear radii.  
 This is why even today, in the investigation of break-up processes 
 of weakly-bound systems close to the drip-line, attention is
 focused in the role of the former and nuclear effects have been often 
 ignored.
\par
 A radically different picture emerges, however, when one considers 
 the most general origin of nuclear transition couplings and forsakes 
 concepts specifically tied to ordinary collective modes.
 If the presence of loosely-bound orbitals pushes out the inelastic
 transition densities into the continuum to radial ranges in the order 
 of 15--25 fm, consequences of this significant shift will 
 {\it inevitably reflect in both the Coulomb and nuclear formfactors}.
 Thus one can anticipate situations for weakly-bound neutron/proton-rich 
 systems where the nuclear couplings may still be dominant under the
 diluted-density conditions that prevail at separation distances of
 the order of 20 fm.
\par
 The possibility of observing such effects is tantalizing and we intend 
 in this contribution to provide concrete estimates for the radial 
 ranges that are likely to be relevant in actual measurements. 
 To this end we consider in what follows the break-up of \b8 into  
 \be7 induced by a high-energy collision with lead nuclei, a case 
 experimentally explored in ref.$^{\,4)}$.  
 Furthermore, to illustrate how critical it can be the extension
 of ordinary ``safe-distance'' concepts into reactions involving 
 weakly-bound systems we shall also examine the same break-up process
 but on low-energy collisions with a nickel target, a reaction discussed
 in ref.$^{\,5)}$.
\par
 In Section 2 we introduce the expressions used to construct the
 matrix elements that induce single-particle proton transitions into 
 the continuum.  
 A brief review of a coupled-channel formalism to evaluate semiclassical
 reaction amplitudes is also included here.
 Results of this calculation scheme in terms of transition
 densities, formfactors, cross sections and Q-value distributions
 are given in Section 3.  
 More specifically, Section 4 is devoted to reactions at energies below
 the barrier.  Examples of angular distributions are shown in this
 context.
 Concluding remarks close the presentation in Section 5.
\vskip 1.5truecm
\pni
{\sect 2. Formalism}
\par
 Microscopic calculations of formfactors for inelastic transitions
 have been routinely performed in the past.  See, for example the review
 article in ref.$^{\,6)}$.
 The basic expressions that are found in the literature need only be adapted 
 to account for the coupling scheme that is exploited in our specific 
 application.
 We adhere here to the familiar prescription$^{\,7)}$~in which the ground 
 state of \b8 is obtained by coupling the valence proton in a
 single-particle $p_{3/2}$ state with an inert \be7 core with 
 quantum numbers $I_c^\pi$=3/2$^-$.
 The experimentally known binding energy ($-$0.14~MeV)
 can then be reproduced by choosing a single particle potential for the 
 protons with a Woods-Saxon shape and parameters $V_\circ$=$-$44.66~MeV, 
 $r_\circ$=1.25~fm, 
 $a_\circ$=0.52~fm and a spin-orbit coupling of $F_{so}$=0.351~MeV. 
 Excited configurations of \b8 follow from the same coupling scheme
 but include a suitable adjustment of the depth of the 
 single-particle potential, $V_\circ$, to reproduce the corresponding 
 binding energies.
 For a table of potential parameters covering these situations cf.
 ref.$^{\,7)}$. 
\par
 We start by writing the projectile wavefunction for the initial 
 state with angular momentum quantum numbers $J_i$, $M_i$ as

$$
 \Psi_{J_iM_i} = \sum_{m_1 m_c} 
  \langle j_1m_1j_cm_c|J_iM_i \rangle
   \psi_{n_1\ell_1j_1m_1} (\vec r,\sigma)
    \phi_{c,j_cm_c} (\vec r_c,\sigma_c)~,
\numeq
$$
\pni
 where $\vec r$, $\sigma$ are the space and spin variables for the
 proton and $\vec r_c$, $\sigma_c$ are the corresponding global
 quantities for the \be7 core.
 The single-particle proton wavefunction $\psi$
 corresponds to a bound state
 and is fully determined by specifying, in addition to $j_1,\,m_1$, the 
 principal quantum number $n_1$ and the orbital angular momentum $\ell_1$.
 Similarly, for the final state we use

$$
 \Psi_{J_fM_f} = \sum_{m_2 m_c'} 
  \langle j_2m_2j_cm_c'|J_fM_f \rangle
   \psi_{E\ell_2j_2m_2} (\vec r,\sigma)
    \phi_{c,j_cm_c'} (\vec r_c,\sigma_c)~,
\numeq
$$
\pni
 where the fundamental difference with (1) is that the single-particle
 proton function corresponds now to a scattering state in the continuum 
 and is thus labeled by the energy $E$.  
 Only the modulus of $\vec k$ enters in the identification of the states, 
 its orientations being incorporated in a proper normalization 
 factor$^{\,8)}$. 
\par
 The matrix elements of the one-body excitation operator 
 $V(|\vec r - \vec R|)$  --~involving only the proton coordinates and 
 the distance $\vec R$ between the reactants~--
 introduce a straightforward integration over $\vec r_c$, $\sigma_c$.  
 This yields a $\delta_{m_cm_c'}$ which, in turn, sets $m_1$=$M_i-m_c$
 and $m_2$=$M_f-m_c$.  Thus, the formfactor for the inelastic transition
 of the proton into the continuum can be written as

$$
 F_{J_iM_i\to J_fM_f} (\vec R) = 
 \sum_{m_c} 
 \langle j_1 (M_i-m_c) j_c m_c| J_i M_i \rangle
 \langle j_2 (M_f-m_c) j_c m_c| J_f M_f \rangle
$$
$$
 \null~~~~~~~~~~f_{n_1\ell_1j_1(M_i-m_c)\to E\ell_2j_2(M_f-m_c)} (\vec R)~.
\numeq
$$
\pni
 Here the function $f$ stands for the basic building block in the 
 microscopic construction of inelastic formfactors, namely the matrix 
 element inducing the transition of the valence proton between two pure 
 single-particle states,

$$
 f_{n_1\ell_1j_1(M_i-m_c)\to E\ell_2j_2(M_f-m_c)} (\vec R) =
$$
$$
 =  \sqrt{\pi}~\sum_{\lambda} 
     (-1)^{M_f-m_c+\half}~
 \delta(\ell_1+\ell_2+\lambda,{\rm even})~
  \langle j_1 \half j_2 -\half | \lambda 0 \rangle~ 
$$
$$
 \times~
 { {\sqrt{2j_1+1}\sqrt{2j_2+1}} \over {\sqrt{2\lambda+1}} }~
 \langle j_1 (m_c-M_i) j_2 (M_f-m_c) | L (M_f-M_i) \rangle~
         Y_{\lambda(M_i-M_f)} (\hat R)
$$
$$
 \times~\left[ \int_{0}^{\infty}r^2 dr \int_{-1}^{+1}~du 
    R^*_{E\ell_2j_2} (r) R_{n_1\ell_1j_1} (r)
      V \big ( \sqrt{r^2+R^2-2rRu} \big ) P_\lambda (u) \right ]~.
\numeq
$$
\par
 We distinguish at this point the nuclear and Coulomb components of the 
 excitation field, $V = V^{^N} + V^{^C}$.  The former 
 involves the mean field felt by the proton
 due to the presence of the target and is generally parameterized
 by a Woods-Saxon function of the form

$$
 V^{^N} \big( |\vec r-\vec R| \big) = 
   { {V_{p_T}} 
        \over 
     {1 + \exp\big[(|\vec r-\vec R|-R_{_T})/a_{p_T})\big]} }~.
\numeq
$$
\pni
 The quantities $V_{p_T}$ and $a_{p_T}$ should not be confused with the 
 potential depth and diffuseness previously invoked (that referred to the 
 projectile mean field).  They assume, nevertheless, close numerical 
 values$^{\,13)}$. 
 The Coulomb component is given by
 
$$
 V^{^C} \big(|\vec r-\vec R| \big) = \Phi_\circ~ 
 { { R_{_T} } \over {|\vec r-\vec R|} }
 ~,~~~~~~~~~~~~~~~~~~{\rm  for}~
 |\vec r-\vec R| \ge R_{_T}
\numeq
$$
$$
 V^{^C} \big( |\vec r-\vec R| \big) = \Phi_\circ~  
 \left[
 { {3} \over {2} } -
 { {|\vec r-\vec R|^2} \over {2R_{_T}^2} }
 \right]
 \,,~~~~~~{\rm  for}~
 |\vec r-\vec R| < R_{_T}
\numeq
$$
\pni
 in terms of $\Phi_\circ = z_{ef}z_{_T}e^2/R_{_T}$.  The effective
 charge is equal to one for all multipolarities except $\lambda$=1, 
 when $z_{ef}$=$N_{_P}/A_{_P}$ ensures that no spurious motion of the
 center of mass takes place.
\par
 The screening effects taken into account by the splitting of the Coulomb
 field into the two expressions (6,7) are often ignored.  Conceptually
 more disturbing is, however, the fact that standard analyses of the Coulomb 
 excitation processes for neutron-rich, weakly-bound systems have 
 consistently relied on the effects of the multipole response to the
 field $r^\lambda$.
 This came about from a rather uncritical use of the expression

$$
 { {1} \over {\sqrt{r^2+R^2-2rRu}} }= \sum_{L} 
 { {r^\lambda} \over {R^{\lambda+1}} }
 P_\lambda(u)~. 
\numeq
$$
\pni
 which allows --~when introduced in (4)~-- to perform the integral
 over $u$ analytically and thus reduce the radial dependence of the 
 Coulomb formfactors to the familiar $R^{-\lambda-1}$ form.  
 The expansion (8) is, however, valid only when $r<R$.  
 This may be a reasonable assumption for Coulomb excitation in
 reactions with stable systems and at low bombarding energies.  
 It is, on the other hand, highly questionable for collisions
 involving weakly-bound systems.  In fact, in this case the initial
 wavefunction extends so much outwards that a major part of 
 the reaction cross sections does originate at distances where
 the target gets to be ``inside'' the transition density. 
\par
 We have set to investigate in this contribution 
 the role of the nuclear couplings into the continuum excitation
 spectrum.
 While the Coulomb field happens to be proportional to a generating 
 function for the Legendre polynomials, no equivalent simplification is 
 available to handle the nuclear field given in eq.~(5).  
 But then, once we are forced by the characteristics of the latter to 
 perform numerically the double-integral in eq.~(4), there is no compelling 
 reason to continue using the (incorrect) $r^\lambda$-response for the 
 Coulomb excitation aspects of the reaction analysis.  
 Actually, by constructing the formfactors for the nuclear and Coulomb 
 components on equal footing one should be able to judge the 
 quality of the approximation introduced in previous treatments.
 Notice that by keeping the Coulomb integrand explicitly as a 
 function of $\sqrt{r^2+R^2-2rRu}$, an appropriate expansion into either 
 $r^\lambda/R^{\lambda+1}$ or $R^\lambda/r^{\lambda+1}$ is automatically
 insured even for the point-field given in eq.~(6).
\par
 It was already stated that the radial wavefunctions $R_{n_1\ell_1j_1} (r)$
 have been constructed by solving the Schroedinger's equation for
 the bound states in a nuclear potential with a Woods-Saxon shape, 
 including Coulomb, centrifugal and spin-orbit components.
 Similarly, the continuum states $R_{E\ell_2j_2} (r)$ have been 
 obtained by matching the asymptotic Whittaker's functions to 
 the numerical solutions constructed within the potentials' range.
\par 
 Finally, we briefly recall the reaction formalism in which the 
 formfactors constructed according to these prescriptions are to be 
 applied.
 We exploit a semiclassical scheme to construct elastic and inelastic 
 cross sections$^{\,9)}$.  
 Asymptotic values of the reaction amplitudes are obtained for each 
 partial wave $\ell$ as a function of time by solving a set of coupled 
 differential equations of the following form

$$
 \dot a _i^{\ell} (t)~ =~ -i\hbar 
 \sum_\beta \, 
 F_{ij}[{\vec R}(t)] \,\exp[-i(\epsilon_i-\epsilon_j)/\hbar]\,
 a_j^{\ell} (t)~,
\numeq
$$
\pni
 where $\epsilon_i$ is the intrinsic excitation of the channel $i$
 and ${\vec R}(t)$ represents the trajectory of relative motion.
 This function is integrated as part of the set of time-dependent 
 equations, assuming an ion-ion potential parameterized according
 to ref.$^{\,11)}$ ($V_\circ$=-42.1~MeV, $R$=9.4~fm and $a$=0.63~fm).
 A first-order elastic phase shift $\delta_{\ell}$
 is also extracted by integrating along the trajectory an optical 
 potential whose imaginary part is assumed to have various 
 plausible forms, as discussed later in the text.
 Details on the discretization procedure for the continuum states 
 and the explicit expressions used to construct differential and total 
 Q-value distributions can be found in ref.$^{\,10)}$.
\vskip 1.5truecm
\pni
{\sect 3. Results for the Reaction \b8 + \pb208 at 46 MeV/nucleon}
\par
 We start by providing some insight on the elements that enter
 in the construction of the formfactors and the origin of the
 energy-dependence of the strength distributions.  We use for this 
 illustration the response associated with the simple $r^\lambda$ 
 radial field.
 The reaction that will be considered throughout this section is
 \b8 + \pb208 at a high bombarding energy of 46 MeV/nucleon (see 
 ref.$^{\,5)}$).
 The top of Fig.~1 shows the radial dependence of the continuum 
 wavefunctions $R_{E\ell_2j_2} (r)$ for $\ell_2$=0 and $j_2$=1/2 
 at three excitation energies, $E$=0.3~MeV (dotted line), $E$=0.6~MeV 
 (solid line) and $E$=1.9~MeV (dashed line).  
 These energies have been chosen to be slightly below, at, and slightly 
 above the energy where the maximum of the response to the 
 chosen field occurs.  Note the rapid change of the 
 wavelength near the continuum threshold which makes possible the optimal
 matching referred to in the Introduction.
 The lower frame displays the radial integrand 
 $R^*_{E\ell_2j_2} (r) R_{n_1\ell_1j_1} (r)\,r^{3}$ for the dipole case,
 $n_1\ell_1j_1\equiv(1,1,1/2)$, $\ell_2j_2\equiv(0,1/2)$
 and for the same three values of $E$.
 One can here appreciate the unusually large portion of 
 space that is involved in the construction of the couplings between
 a weakly-bound state and the continuum.  For a reference consider
 the shaded area, where transition densities corresponding to
 ordinary well-bound states in a system of the size of boron would be 
 entirely localized.
 Fig.~2 shows the energy distribution of the
 electromagnetic dipole strength, $dB(E1)/dE$, for the
 transitions $p_{3/2} \to s_{1/2}$ and $p_{3/2} \to d_{5/2}$ 
 (top and bottom frames, respectively).  In both cases,
 the results obtained for the proton excitation (solid lines)
 are compared with those obtained for the corresponding 
 neutron excitation, under the assumption of an equal
 binding energy for the initial $p_{3/2}$ state (dashed lines). 
\par
 The results collected in Figs.~1,2 are qualitatively similar
 to the ones reported in ref.$^{\,2)}$~ for the case of uncharged particles.  
 Note, however,
 that the calculations in that reference pertained single-neutron 
 transitions in heavy nuclei and were obtained by exploiting
 analytical expressions valid only for square-well potentials.  
 In addition to the effects of a finite diffuseness, the calculations 
 reported here incorporate the proton confining barrier.  
 This is relevant for the lower part of the energy response.
 Because of the shorter extent of penetration into the classically 
 forbidden region, the strength distributions for proton transitions 
 are, {\it ceteris paribus}, peaked at higher excitation energies 
 and acquire a larger width.
\par
 Examples of the dipole couplings involved in the single-proton 
 transitions $p_{3/2}\to s_{1/2}$ and $p_{3/2}\to d_{5/2}$
 are given in Fig.~3 (left and right sides of the figure, respectively).  
 The radial dependence of the formfactors depicted in frames a) and d) 
 are in arbitrary units as the motivation here is only to compare 
 the magnitude of the Coulomb and nuclear terms.   
 This can be done at the level of the double-integrals in 
 eq.~(4), where the two components blend at the proper 
 relative scale, thus allowing us to ignore angular-momentum algebra and 
 other common factors.
 It is seen that for either single-particle transition, the nuclear coupling 
 dominates at center-of-mass distances from 25 fm inwards.  This is
 significant, given the fact that a considerable part of the break-up 
 cross section originates from partial waves that probe distances inside 
 this domain
\footnote {$^\dagger$} 
{\ninerm \baselineskip 9pt
 Note that for energies of the order of 40 MeV/nucleon a classical
 picture of the relative motion corresponds to one of almost 
 straight-line trajectories and thus the impact parameters and distances
 of closest approach are almost identical.
\smallskip}.
 The dotted lines correspond to the extrapolation into the shorter
 distances of the simple $R^{-\lambda-1}$-dependence.  We can see that
 even for the dipole case, significant deviations from the proper answer
 start to register inside 20 fm, building up to factors of about two
 for $R$=10 fm.  This is not a small correction for these partial
 waves, especially if one keeps in mind that cross sections 
 reflect the square of the coupling matrix elements.  It may not
 be overall a major consideration, however, since for $\lambda$=1 a large
 percentage of the total contribution to Coulomb break-up comes
 from the large partial waves.     
 Frames b) and e) display another aspect of the
 formfactors, namely their dependence as a function of the
 continuum energy at a fixed distance.  This has been chosen
 --~for the calculations displayed in the figure~-- to be 15 fm.
 Because of the inclusion of the nuclear component and the marked 
 deviations in the Coulomb term an altogether different population 
 pattern for the energy channels above threshold should emerge.
 This is indeed the case, as it can be seen in the two bottom frames, 
 where the electric response in terms of the square of the matrix elements 
 of the field  $r^\lambda$ is displayed.  Notice the significant
 shift of the maximum of the couplings towards a lower value of the
 energy.
 This change would reflect directly in the Q-value distribution
 of the cross section.  In fact, in the adiabatic limit the profiles of
 both functions coincide. 
\par
 We have not displayed in Fig.~3 the corresponding information for the 
 transition $p_{3/2}\to d_{3/2}$.  Although the spin-orbit coupling
 is properly taken into account in our formalism, there are no qualitative 
 aspects in this transition that differ enough from the $d_{5/2}$ case 
 that merit its inclusion here.  Insofar as calculating cross sections
 is concerned, however, dipole transitions $p_{3/2}\to d_{3/2}$ 
 represent a viable alternative and need to be incorporated in the 
 coupled-channel analysis.
\par
 Fig.~4 is analogous in structure to Fig.~3, except that it covers the
 $\lambda$=3 component of the multipole expansion of the couplings in eq. (4).
 This is not relevant for the $p_{3/2}\to s_{1/2}$ transition
 but is allowed by parity and angular momentum selection rules
 for the $p_{3/2}\to d_{3/2}$ and $p_{3/2}\to d_{5/2}$ transitions
 into the continuum.
 It is known that the dependence in multipolarity of the nuclear 
 formfactors is weaker than in the Coulomb case.
 This is reflected in the figure (referring to the 
 $p_{3/2}\to d_{5/2}$ transition), where an even more pronounced 
 dominance of the nuclear component is put in evidence.
 It is interesting to note how much larger are in the octupole case
 the differences introduced by using the $r^\lambda$-response,
 now associated with a pure $R^{-4}$ radial dependence.  
 Overestimation of the couplings can in this instance involve orders 
 of magnitude.
 By the same token, calculations of cross sections for Coulomb break-up
 relying on the $B(E3)$ strength distribution would lead to totally 
 erroneous predictions (compare frames b) and c)).
 One should stress here that this problem is unrelated to the need
 of including nuclear processes.  That is, the standard procedure to
 calculate electric dissociation for weakly-bound systems is incorrect 
 even if a justification to leave the nuclear field out of the picture 
 could be found.
\par
 We devote the next two figures to indicate how the full range of
 partial waves contributes to the reaction cross sections and to 
 learn about the effects of an absorptive potential.
 Fig.~5 shows the relative importance of the different impact 
 parameters, $b$, in building the cross section for projectile 
 dissociation.  For a given channel $k$ the quantity $d\sigma_k/db$ 
 incorporates three $b$-dependent factors, namely

$$
 { {d\sigma_k} \over {db} }
 \propto
 b~|a_k(b)|^2
 \exp \left[ -2\int W\big(R(t)\big)\Big|_b\,dt \right]~.
\numeq
$$
\pni
 The modulus squared of the reaction amplitude $a_k(b)$ gives the 
 probability to populate the channel and the exponential introduces,
 within our model space, the attenuation which takes into account 
 all other degrees of freedom not explicitly included in the 
 formalism (for instance, those related to fusion).
 This factor is constructed by integrating along the trajectory
 the imaginary part of the optical potential, $W\big( R(t)\big)$.
 For the calculations shown in this figure the absorptive potential
 has been given the same geometry of the real part and half its
 strength (i.e. $W_\circ$=$-$21~MeV, $R$=9.4~fm, $a$=0.63~fm). 
 We focus our attention on a continuum channel at energy $E$=0.6~MeV
 and consider for the moment only the dipole transition between the
 $p_{3/2}$ and $s_{1/2}$ states.
\par
 The coupled-channel program has been run three times: with Coulomb
 couplings only, with nuclear couplings only and with both types
 of interactions
 included.  The corresponding results for $d\sigma_k/db$ are shown
 with dotted-, dashed- and solid-lines, respectively.
 One can see that the nuclear component builds its contribution
 mostly from impact parameters between 10 and 20~fm.
 In this range it dominates over the Coulomb part.  This one, on
 the other hand, extends much farther out, as it can be expected 
 from the slow $R^{-2}$ radial dependence of the couplings.
 The two components --~whose formfactors at large distances differ 
 in sign~-- can interfere in a significant way.  Notice, for instance,
 the dip in $d\sigma_k/db$ at $b\approx$18~fm where the separate 
 terms have almost identical magnitude, in close correlation with 
 the crossing of the formfactors already shown in Fig.~3.
 From this figure one can infer that it is safe to ignore nuclear
 effects for impact parameters larger than $b\approx 30$~fm.
 This value of $b$ corresponds to an elastic Coulomb scattering angle
 of about three degrees.
\par
 A characteristic ratio of one-half between the imaginary and real parts 
 of the optical potential is often used in the heavy-ion literature in 
 the lack of better judgment.  We note, however, that changing this
 proportion even by factors of two does not affect the qualitative aspects 
 of what has been displayed in Fig.~5.
 This is due to the fact that the range of the absorption is what 
 mostly determines the effective cut-off for central impact parameters.
 In a situation involving a projectile with a pronounced proton skin
 it seems appropriate to test, in addition, the sensitivity of the results 
 to changes in the absorption range.
 This question is addressed in Fig.~6, where the radius of the imaginary 
 part of the potential used in the previous figure has been shifted by 
 $\pm$1~fm.  The scope is here slightly different from the one in the
 previous application, as we plot the function $d\sigma/db$ not for 
 an individual channel but integrated over the entire Q-value range,

$$
 { {d\sigma} \over {db} }
 \propto
 b~[1-|a_\circ|^2] 
 \exp \left[ -2\int W\big(R(t)\big)\,dt \right]~.
\numeq
$$
\pni
 In this expression the first quantity between brackets gives now
 the global probability to be taken away from the elastic channel.
 The nuclear component (building up its contribution closer to the 
 central collisions) is naturally more sensitive to the modulation 
 in range than the Coulomb part.
 The predominant role of the nuclear processes in the 10--25~fm 
 interval of impact parameters is not, however, affected.  
 Note that the scales of the nuclear frames are, for both dipole 
 and octupole transitions, one order of magnitude larger than the 
 corresponding ones for the Coulomb case.
 It is also interesting to observe that, as a consequence of the 
 combination of many positive-energy channels, the distributions 
 become rather structureless.  
\par
 Examples of the Q-value dependence of the cross sections for 
 transitions into the continuum are shown in Fig.~7.  To the left of 
 the figure we consider the single dipole transition 
 $p_{3/2}\to s_{1/2}$ and present the separate Coulomb and 
 nuclear components together with the result of their combined action. 
 The interfering character of the two mechanisms is here quite evident.  
 Thus, even though the nuclear contribution is {\it per se} larger than 
 the Coulomb one, the total cross section is only moderately increased 
 with respect to what would have been (wrongly) predicted in terms of 
 Coulomb dissociation only.
 There is a small shift of the distribution to higher Q-values and the 
 distribution becomes wider as well.
 The right-hand-side of the picture displays the total cross section
 obtained for the transitions $p_{3/2}\to s_{1/2}$, $p_{3/2}\to d_{3/2}$
 and $p_{3/2}\to d_{5/2}$ including the dipole and octupole terms,
 i.e. all the multipolarities allowed by parity and angular momentum
 selection rules.  Notice that the addition of the octupole component
 increases the tendency to populate channels with higher excitation 
 energy.  
\par 
 In connection with Figs.~3,4 attention was brought to the strong
 deviations from the true Coulomb formfactors that could be introduced 
 by an indiscriminate reliance on the $r^\lambda$-response.
 We show in Fig.~8 two concrete examples of the actual manifestation 
 of this effect in reaction cross sections.  As an illustration
 we have here calculated Q-value distributions for the specific 
 transition $p_{3/2}\to d_{5/2}$;  
 the continuum $d$-channels are accessible via both dipole and 
 octupole couplings and thus provide a good case to explore the 
 dependence with multipolarity.  
 The solid lines correspond to calculations performed with 
 the microscopically-constructed Coulomb formfactors while the 
 dashed curves have been obtained by replacing them by
 others with the simple $R^{-\lambda-1}$ radial dependence.
 While results for the dipole case are not affected in a major way 
 by the substitution --~and then only for the lower excitation 
 energies~-- an improper handling of the octupole couplings would 
 lead to radically different answers.  The cross sections are grossly 
 overestimated (as it had been anticipated) but even the profile of 
 the excitation spectrum is altered.
\par
 We close this section devoted to the \b8 + \pb208 reaction at
 high bombarding energies with Fig.~9.  We show here a 
 break-down into components of the Q-value distribution for all 
 the inelastic transitions starting from the bound-, $p_{3/2}$, 
 state in \b8 that involve the $s$ and $d$ states in the continuum.
 The total result is given by the thick solid line and displays a 
 maximum at energies just below the one-MeV level (see also Fig.~7).  
 The other curves show the same quantity, but now only some specific
 channels and/or multipolarities were included in the 
 calculations.  Setting aside the continuum $j$-quantum number (whose 
 values are left implicit) we identify the different situations as follows:
 dipole $p \to s$ transitions (dashed line), dipole $p \to d$ transitions 
 (dotted line), octupole $p \to d$ transitions (dash-dotted line) and 
 dipole-plus-octupole $p \to d$ transitions (thin solid line).
 It can now be inferred that the fact that the distribution peaks 
 at a low energy is due to a relative predominance of the 
 dipole $p\to s$ transition in that part of the spectrum.
 Note, however, that due to the coupled-channel character of the 
 calculation the additivity of these partial results is not entirely
 justified.
 Although the order of magnitude of the cross section seems to be correct,
 at this level the profile of the $Q$-value distribution
 deviates from the scarce available evidence. Of course any attempt to 
 compare with data should not ignore
 electric and magnetic transitions of other multipolarities,
 which have been discussed in refs.$^{\,7)}$.
\vskip 0.9truecm
\pni
{\sect 4. Low-energy Case}
\par
 We have already emphasized how the long-range of the nuclear 
 formfactors for inelastic excitation of weakly-bound 
 single-particle states leads to features which are markedly
 different from those encountered in the excitation of stable 
 systems.  
 We turn in this section to a class of reactions that are normally 
 considered to be totally dominated by electromagnetic processes, 
 namely collisions at bombarding energies below the Coulomb 
 barrier.  
 Within a classical picture one argues that at these energies the 
 trajectories of relative motion cannot explore the internal region 
 where nuclear forces are effective and that, consequently, excitation 
 processes via the nuclear field should be negligible.
 The possibility of having a ``clean'' Coulomb dissociation process was, 
 in fact, one of the main motivations for the experiment presented 
 in ref.$^{\,5)}$, a study of the break-up of $^{8}$B on 
 a $^{58}$Ni target at a subbarrier bombarding energy of  $E$=25.3 MeV.  
 At this energy, even for the most penetrating, head-on, collisions the 
 distance of closest approach is of the order of 10~fm.  This is
 several nuclear diffuseness outside of the sum of the radii of the 
 reactants and therefore regarded as ``safely'' outside of the
 nuclear range of action.
\par
 In order to check this point, we have performed calculations of inelastic
 excitation to the continuum for the $^{8}$B + $^{58}$Ni reaction at the
 same bombarding energy of the experiment in ref.$^{\,5)}$.  
 Among the contributing processes with different multipolarity we
 have specifically considered the transition in which the $p_{3/2}$ proton
 in $^{8}$B is promoted, via dipole excitation, to a $s_{1/2}$-state 
 in the continuum.  
 The angular distribution for the global inelastic process
 --~i.e. integrated over all possible excitation energies~--
 is given in Fig.~10.
 We again compare the full calculation (resulting from the total
 proton-target interaction) with those that only include either 
 the Coulomb or nuclear couplings.
 As in a standard situation, the distribution at forward-angles
 is dominated by the Coulomb excitation mechanism.
 Nuclear processes, on the other hand, prevail at more backward 
 angles, with a characteristic grazing peak followed by a decrease 
 of the cross sections due to absorption.  
 Note, however, that the peak in the angular distribution 
 (at $\theta\approx\,$80$^{\circ}$) is associated with a classical trajectory 
 whose distance of closest approach is 12~fm, i.e. considerably larger 
 that the expected range of nuclear forces.  
 Even the trajectory leading to the final angle $\theta$=40$^{\circ}$, 
 for which Coulomb and nuclear processes contribute just about equally, 
 has a distance of closest approach of about 20~fm!
\par
 The relative importance of the Coulomb and nuclear contributions 
 (which changes with the scattering angle) should also show up in 
 different $Q$-value profiles.  
 As an example, we display in Fig.~11 the double
 differential cross section as a function of the excitation energy at
 fixed values of the scattering angle.  The three curves in each frame
 refer to the full calculation (resulting from the total
 interaction) and to calculations that only include either 
 the Coulomb or nuclear couplings.  
 As expected, the forward-angle region (top frame) is dominated by 
 Coulomb dissociation and the shape of the $Q$-distribution
 approaches that of the $B(E1)$ distribution (cf. Fig.~3c).  
 This correspondence becomes even closer for smaller angles, as the 
 relevant trajectories probe the formfactors in more external regions.
 The situation is different for the large angles, where the nuclear 
 contribution is dominant (bottom frame).  
 The $Q$-value distribution is now peaked at a rather higher energy, 
 reflecting the energy dependence of the formfactors at smaller radii 
(cf. Fig.~3b). 
 Destructive interference between the nuclear and Coulomb contributions 
 leads to a practically vanishing cross section at an intermediate angle, 
 $\theta$=40$^\circ$ (middle frame).
\vskip 1.5truecm
\pni
{\sect 5. Closing Remarks}
\par
 We aimed in this paper to study the role of Coulomb
 and nuclear processes in the dissociation of \b8.  We have 
 deliberately refrained from developing a new reaction formalism 
 and chosen instead to approach the problem with well-known tools 
 introduced for the treatment of inelastic processes in heavy ion 
 reactions.
 These call for the formfactors to be constructed from a folding
 of the relevant transition densities with the Coulomb and
 nuclear excitation fields.  The procedure, by construction, treats
 both components on equal footing and immediately provides ways to 
 compare the relative importance of the coupling strengths.
\par
 From this point of view it is quite clear that if a given mechanism 
 manages to stretch the transition densities out --~as it is the case 
 when weakly-bound orbitals come into play~-- the situation is likely 
 to affect the nuclear component much more than the Coulomb one.  
 In fact, the Coulomb coupling, being naturally long-ranged, cannot 
 benefit from this novel feature as much as its nuclear counterpart. 
 The latter, as it is well-known, does rely on the close contact 
 between densities and excitation fields.
\par
 The conclusions we have reached concerning the primary role of the 
 nuclear processes in the proton dissociation of \b8 are significantly
 at odds with the standard interpretation of this phenomenon. 
 Clearly it does not make sense to extract from the available data
 any conclusion on interfering E1, M1 and E2 processes if the 
 electromagnetic couplings do not play a dominant role.
 In this perspective reported fittings to data, when only based 
 on Coulomb dissociation, may have to be dismissed altogether.  
 Actually, a critical inspection of Coulomb dissociation studies could 
 have provided some warning about their weakness, since the range of 
 relevant impact parameters was often loosely specified or merely chosen 
 to reproduce the observed cross sections.
 We have let, on the other hand, ordinary nuclear dynamics dictate
 on this question and the answers --~even allowing for rather different 
 absorption geometries~-- consistently reveal a substantial range of 
 partial waves where the nuclear couplings dominate.
\par
 One could argue that our formalism, developed and successfully tested 
 for heavier ions, may need some adjustment if transported to the realm 
 of light, weakly-bound nuclear systems.
 Still, the magnitudes presently at play so significantly tilt the 
 balance of the reaction in the direction of the nuclear component 
 that the resulting patterns should survive even more than just a fine 
 tuning of the reaction mechanism.
\par
 To shed additional light on these fundamental questions consideration 
 should perhaps be given to an alternative formulation.  We refer to 
 that of treating the dissociation of the projectile in terms of the 
 transfer of the proton to the target in either bound or continuum 
 states.  A good understanding of this complementary view could 
 help revealing eventual shortcomings in the approach that 
 we have followed to reach our conclusions.  We understand that
 investigations along this track are being pursued$^{\,12)}$.
\par
\vskip 1.7truecm
\pni
 This work was partially supported by the Danish Natural Science Research 
 Council, the Danish Ministry of Education and by the INFN.
\vfill
\eject
\null
\vskip 1.4truecm
\baselineskip 13pt
\parskip 0.1truecm
\pni
 {\sect References}
\vskip 0.5truecm
\item{1.}  H. Sagawa, N. Van Giai, N. Takigawa, M. Ishihara and K. Yazaki,
 in Proc. Riken Int. Workshop 1993; H. Sagawa, N. Van Giai, N. Takigawa,
 M. Ishihara and K. Yazaki, Z. Phys. {\bf A351} (1995) 385 
\item{2.} F. Catara, C.H. Dasso and A. Vitturi, Nucl. Phys. {\bf A602}
 (1996) 181
\item{3.} I. Hamamoto, H. Sagawa and X.Z. Zhang, Phys. Rev. {\bf C53}
 (1996) 765; I. Hamamoto and H. Sagawa, Phys. Rev. {\bf C53} (1996) R1492;
 I. Hamamoto and H. Sagawa, Phys. Rev. {\bf C54} (1996) 2369;
 I. Hamamoto and H. Sagawa, Phys. Lett. {\bf B394} (1997) 1;
 I. Hamamoto, H. Sagawa and X.Z. Zhang, Phys. Rev. {\bf C55}
 (1997) 2361;
 \hfill\break
 F. Ghielmetti, G. Col\`o, E. Vigezzi, P.F. Bortignon and R.A. Broglia, 
 Phys. Rev. {\bf C54} (1996) 2134R
\item{4.} T. Motobayashi \etal, Phys. Rev. Lett. {\bf 73} (1994) 2680;
 T. Kikuchi \etal, Phys. Lett. {\bf 391B} (1997) 261
\item{5.} J.J. Kolata \etal, Nucl. Phys. {\bf A616} (1997) 137c
\item{6.} R.A. Broglia, C.H. Dasso, G. Pollarolo and A. Winther,
  Phys. Rep. {\bf 48C} (1978) 351.
\item{7.} H. Esbensen and G.F. Bertsch, Nucl Phys. {\bf A600} (1996) 37;
\hfill\break C.A. Bertulani, Nucl. Phys. {\bf A587} (1995) 318; Zeit.
 Phys. {\bf A396} (1996) 293
\item{8.} G.F. Bertsch and H. Esbensen, Ann. Phys. {\bf 209} (1991) 327
\item{9.} K. Alder and A. Winther, {\it Electromagnetic excitations}
 (North Holland, Amsterdam, 1975); \hfill\break
  R.A. Broglia and A. Winther, {\it Heavy ion reactions}
 (Addison-Wesley, Redwood City, 1991)
\item{10.} C.H. Dasso, S.M. Lenzi and A. Vitturi, Nucl. Phys. {\bf A611}
 (1996) 124
\item{11.} O. Akyuz and A. Winther, in {\it Proceedings of the Enrico Fermi
 International School of Physics, 1979, Course on Nuclear Structure
 and Heavy Ion Reactions}, Ed. R.A. Broglia, C.H. Dasso and R.A. Ricci
 (North Holland, Amsterdam, 1991)
\item{12.} A. Bonaccorso, Phys. Rev. {\bf C53} (1996) 849;
 A.Bonaccorso and N. Vinh Mau, Nucl. Phys. {\bf A615} (1997) 245
\item{13.} A. Bohr and B. Mottelson, in {\it Nuclear Structure} Vol. I,
 W.A. Benjamin, Reading, 1969           
\eject
\null
\parindent=1.8truecm
\baselineskip 15pt
\pni
 {\sect Figure captions}
\vskip 0.3truecm
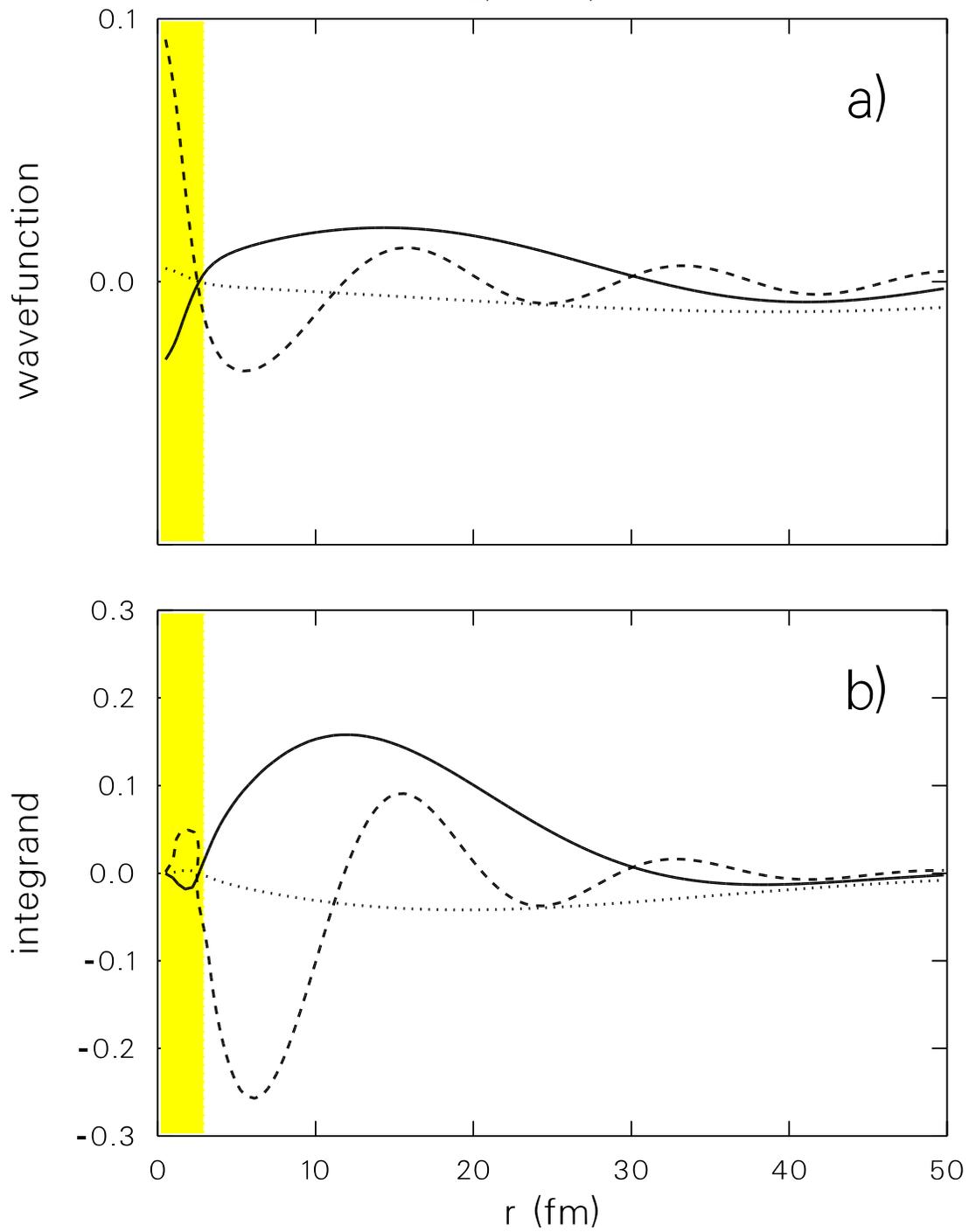
\item{Fig.~1 --}
 a) radial dependence of the $\ell$=0 proton wavefunction in the 
 continuum at three different energies.  These have been chosen to be at 
 the maximum of the $B(E1)$ strength distribution ($E$=0.6 MeV, solid line)
 and two other values, above ($E$=1.9 MeV, dashed line) and
 below ($E$=0.3 MeV, dotted line) this one.
 The shaded area indicates the extent of the binding potential
 in the \b8 nucleus.~~ 
 b) Radial dependence of the integrand which yields the values
 of the $B(E1)$ $r^\lambda$-response for the three energies quoted above.
 The identification of the curves corresponds also to the one for
 the upper frame.  The wavefunction for the initial,
 bound configuration, has been chosen as a $\ell$=1 proton wavefunction
 with a binding energy of 140 KeV. 
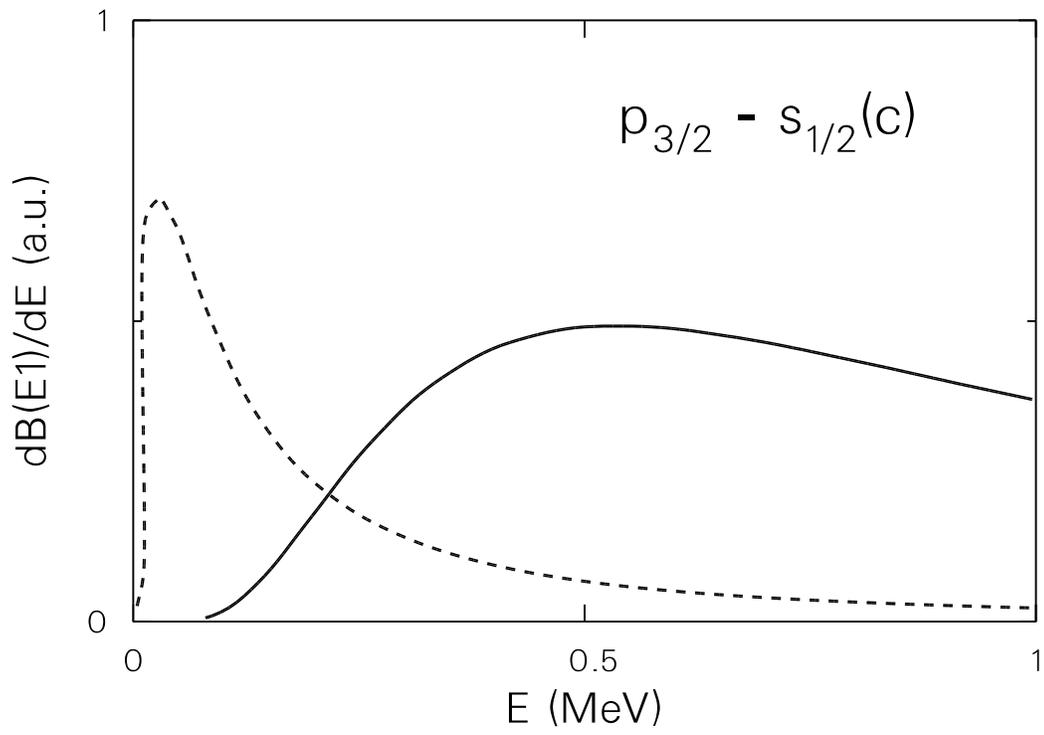
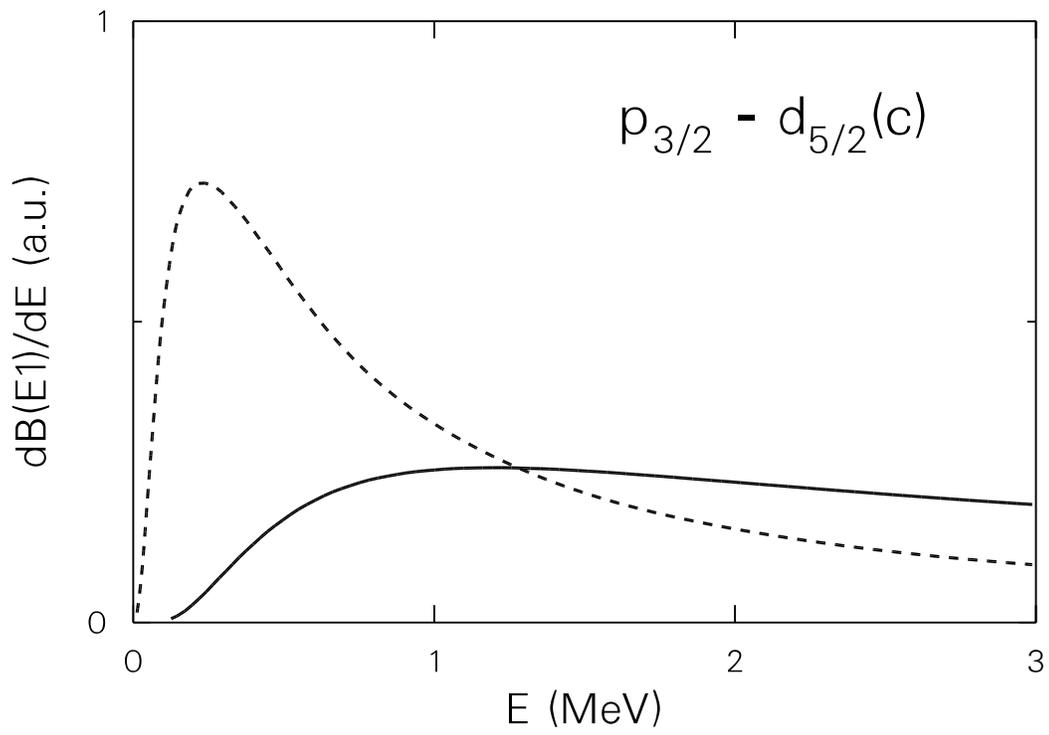
\item{Fig.~2 --}
 Differential distribution of the
 electromagnetic dipole strength, $dB(E1)/dE$, for the
 transitions $p_{3/2} \to s_{1/2}$ and $p_{3/2} \to d_{5/2}$ 
 (top and bottom frames, respectively).  In both cases,
 the results obtained for the proton excitation (solid lines)
 are compared with those obtained for the corresponding
 neutron excitation (dashed lines).  The proton and neutron potentials
 have been adjusted in order to give the same binding
 energy (0.14 MeV) to the initial $p_{3/2}$ state. 
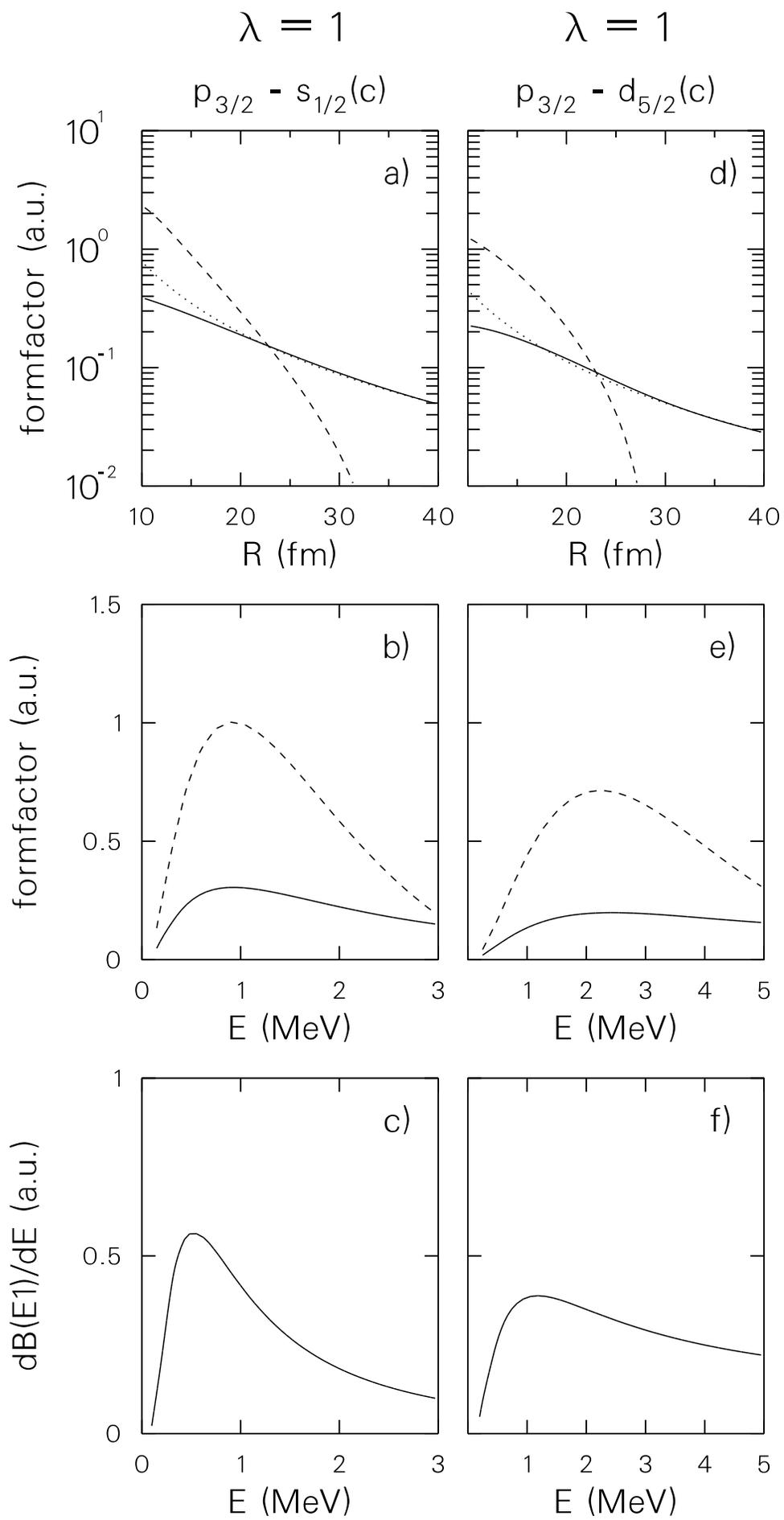
\item{Fig.~3 --}
 Dipole formfactors for the single proton transitions 
  $p_{3/2} \to s_{1/2}$ and $p_{3/2} \to d_{5/2}$ (left and right sides
 of the figure).
 The reaction is \b8 + \pb208 at a bombarding energy of 372 MeV.  
 In the top frames Coulomb and nuclear formfactors
 (solid and dashed lines, respectively) are shown as a function of 
 the ion-ion distance $R$ for an excitation energy in the continuum that
 corresponds to the maximum of the corresponding electric $B(E1)$ distribution
 (shown in the bottom frames). 
 Also shown as dotted lines the extrapolation into shorter distances
 of the simple $R^{-\lambda -1}$ dependence.
 In the middle frames the Coulomb and
 nuclear formfactors are shown as a function of the excitation energy
 for fixed distance ($R = 15 fm$).   For the absolute scales, cf. text.
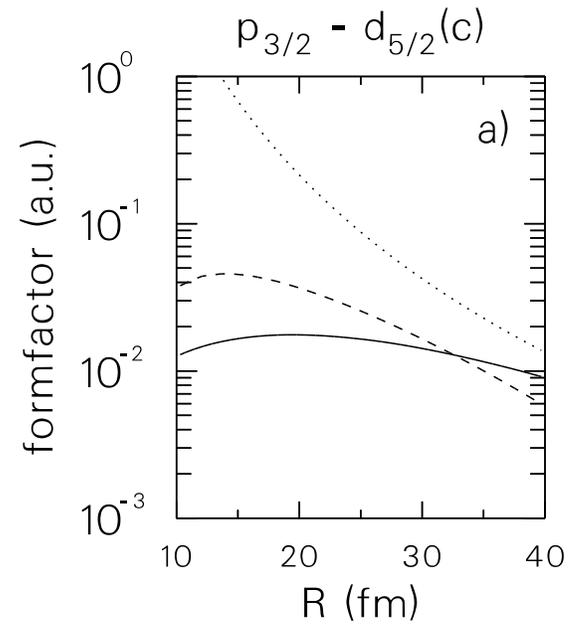
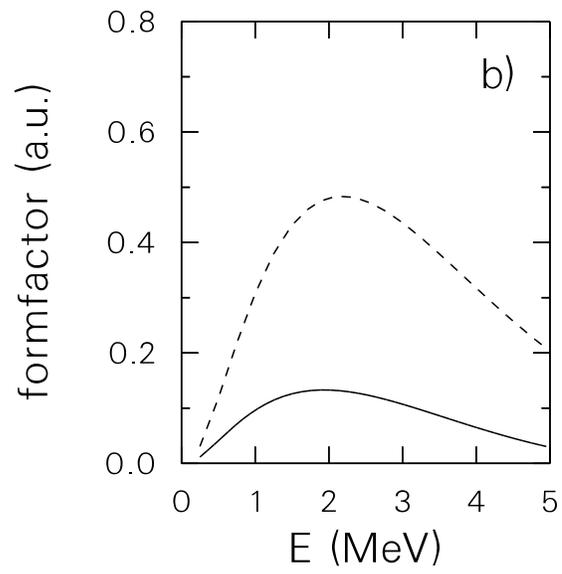
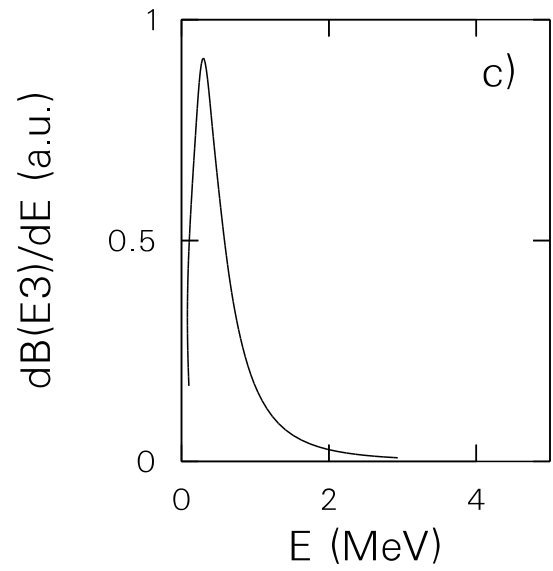
\item{Fig.~4 --}
 Analogous to Fig.~3, but for the transition $p_{3/2} \to d_{5/2}$
 and angular momentum transfer of $\lambda = 3$.  
 Labeling of the curves follows the same convention as in the 
 caption to the previous figure.
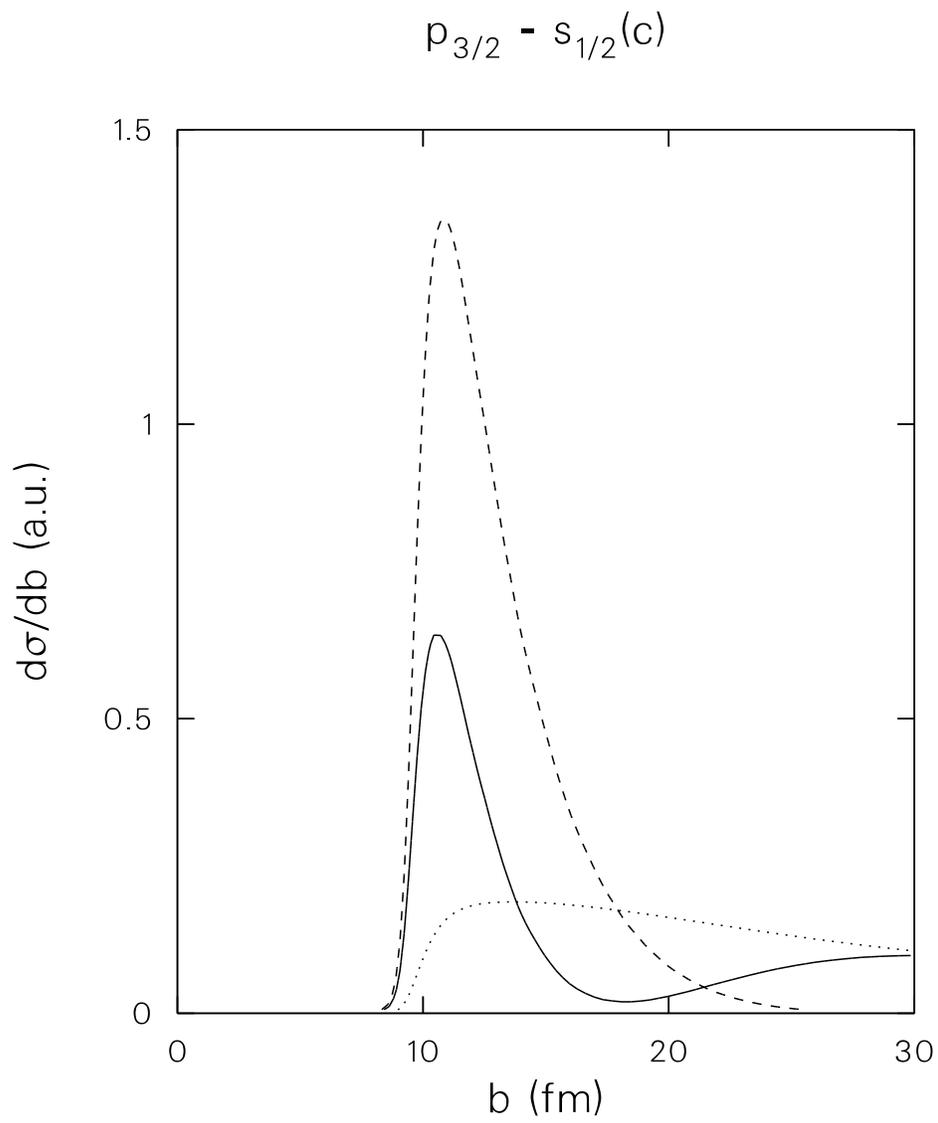
\item{Fig.~5}
 Localization of the reaction cross section as a function of the 
 impact parameter.  
 The reaction is \b8 + \pb208 at a bombarding energy of 372 MeV.
 The calculations are for the single dipole
 transition $p_{3/2} \to s_{1/2}$ and correspond to a channel
 at excitation energy $E$=0.6 MeV.  Three separate runs of the
 coupled-channel program have produced the curves.  They were done
 including Coulomb couplings only (dotted line), nuclear couplings 
 only (dashed line) and both types of interactions (full line).
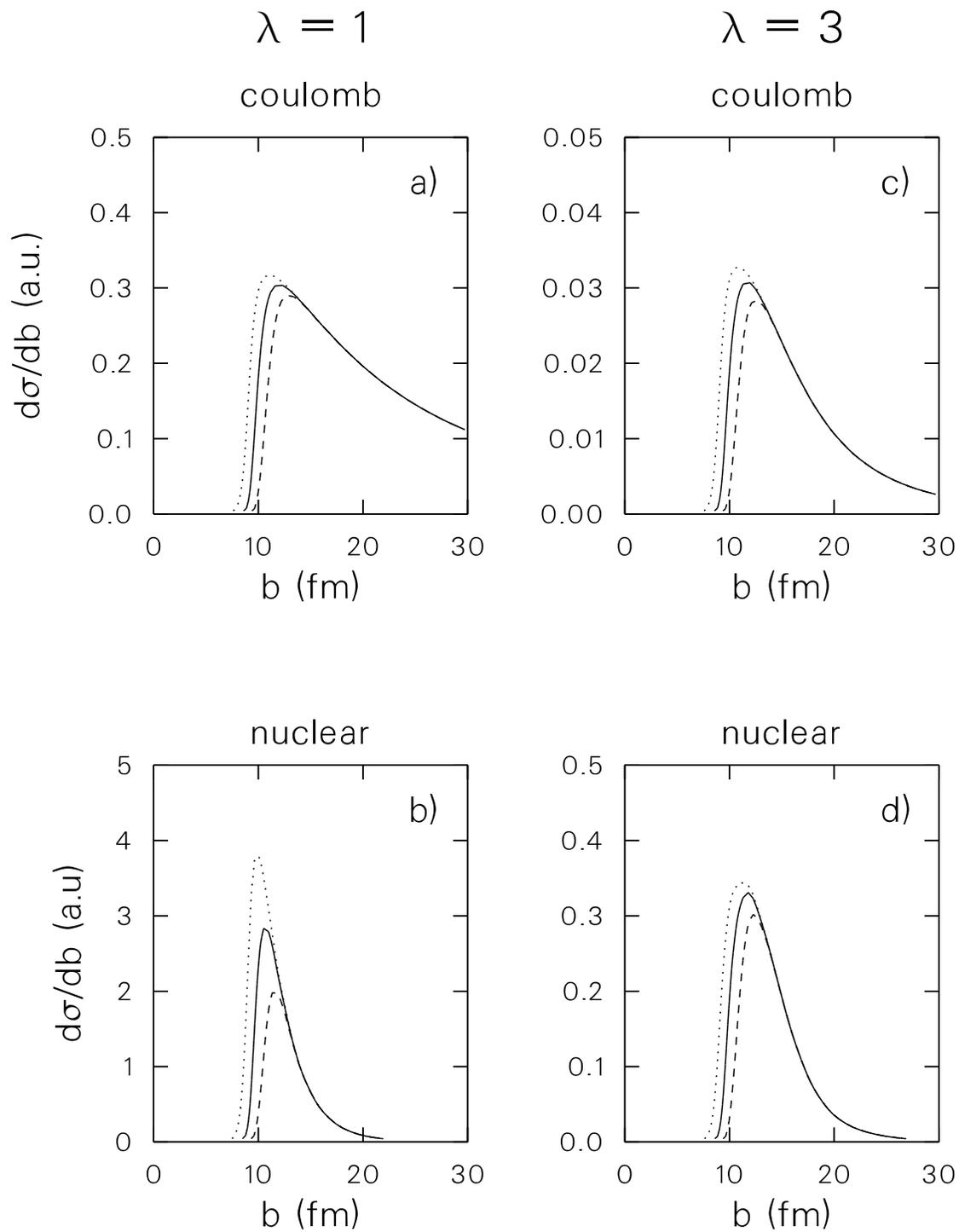
\item{Fig.~6}
 Localization of the reaction cross section as a function of the 
 impact parameter and its dependence as a function of the range
 of the absorptive potential.
 The reaction is \b8 + \pb208 at a bombarding energy of 372 MeV.
 The cross sections depicted here are summed over all the continuum
 channels included for the transitions $p_{3/2} \to s_{1/2}$,
 $p_{3/2} \to d_{3/2}$ and $p_{3/2} \to d_{5/2}$.
 The left column shows the dipole contribution to the multipole
 expansion and the right one the octupole term.
 The full line results from using an imaginary potential with the
 same geometry as the real part and half its strength.  To obtain
 the dashed and dotted curves the radius of the absorptive potential 
 has been changed by adding and subtracting 1~fm, respectively.
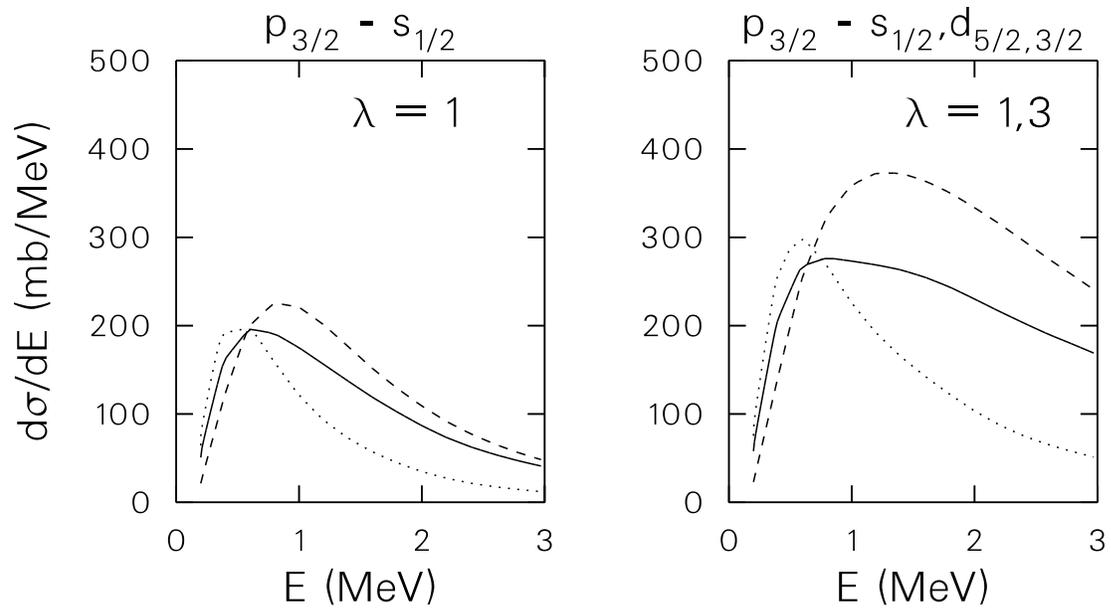
\item{Fig.~7}
 Q-value dependence of the inelastic cross sections into the 
 continuum.
 The reaction is \b8 + \pb208 at a bombarding energy of 372 MeV.
 The curves depicted here emerge from separate calculations
 where the reaction amplitudes have been constructed including  
 Coulomb couplings only (dotted line), nuclear couplings 
 only (dashed line) and both types of interactions (full line).
 The left frame corresponds to the dipole transitions 
 $p_{3/2} \to s_{1/2}$. 
 To the right, the information applies to the total cross section
 accumulated from the dipole and octupole terms and thus exhaust 
 the possibilities afforded by $p \to s$ and $p \to d $ transitions.
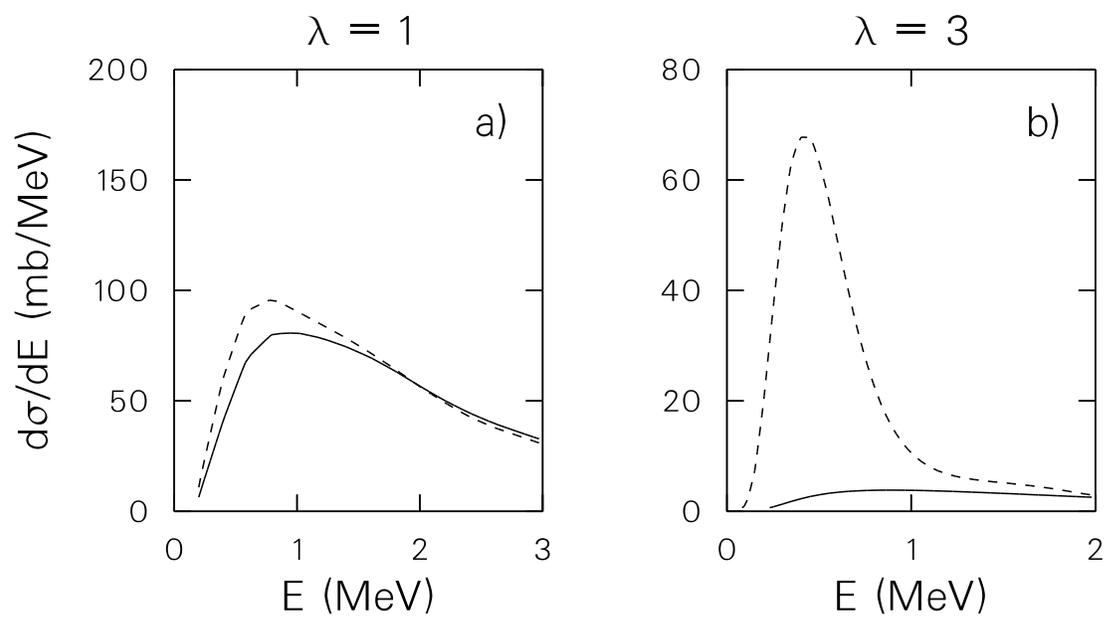
\item{Fig.~8}
 Comparison between cross sections for electric dissociation 
 constructed with the microscopically calculated Coulomb formfactors
 and those with the simple $R^{-\lambda-1}$ radial dependence.
 The frame to the left shows the difference in the predictions for
 the dipole case, while the right frame corresponds to the octupole case. 
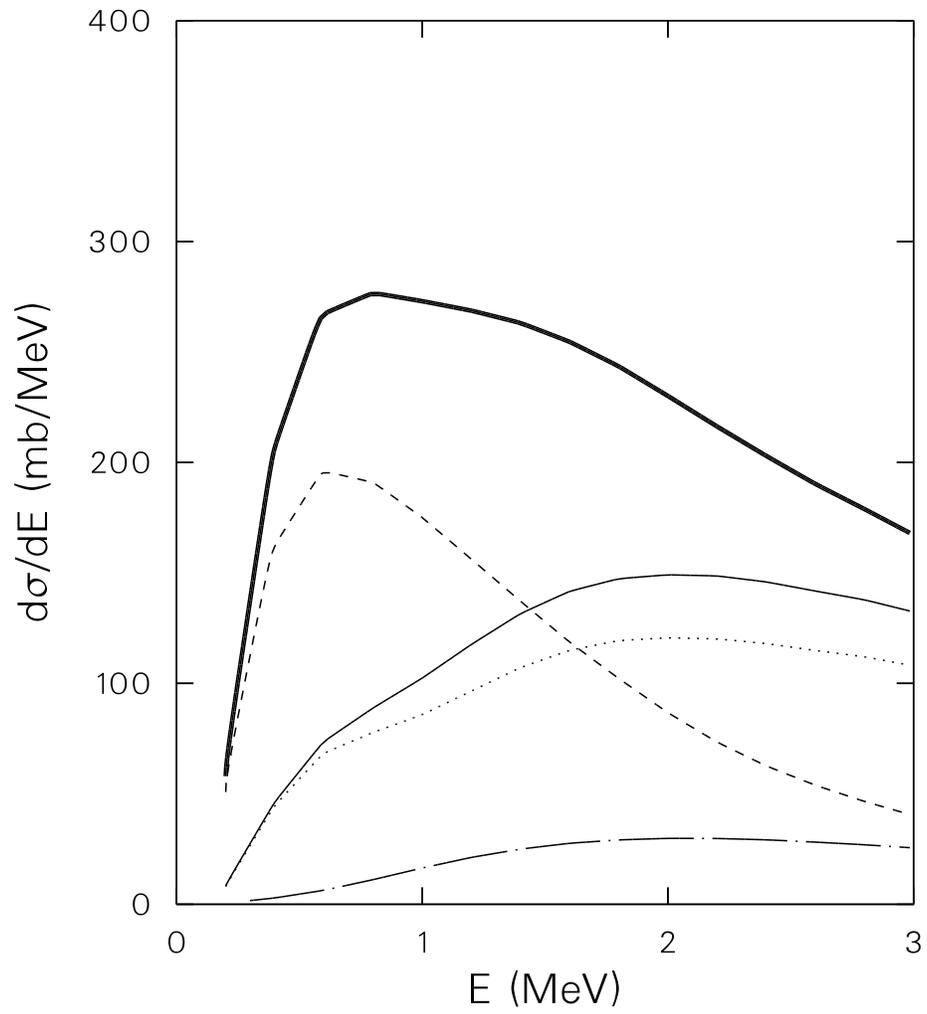
\item{Fig.~9}
 Break-down of the Q-value distribution for all inelastic transitions
 from the bound-, $p_{3/2}$, state in \b8 to $s_{1/2}$ and 
 $d_{3/2,5/2}$ states in the continuum.
 The thick solid line contains the combined effect of the 
 dipole and octupole couplings and thus coincides with the one
 already displayed in Fig.~7.
 The other curves indicate the corresponding quantity but for
 different truncations of this set of reactions channels.
 These are: 
 $p_{3/2} \to s_{1/2}$, $\lambda$=1 (dashed line),
 $p_{3/2} \to d_{3/2}$ and  $p_{3/2} \to d_{5/2}$, $\lambda$=1 (dotted line)
 $p_{3/2} \to d_{3/2}$ and  $p_{3/2} \to d_{5/2}$, $\lambda$=3 (dash-dotted line)
 $p_{3/2} \to d_{3/2}$ and  $p_{3/2} \to d_{5/2}$, $\lambda$=1,3 (thin solid line).
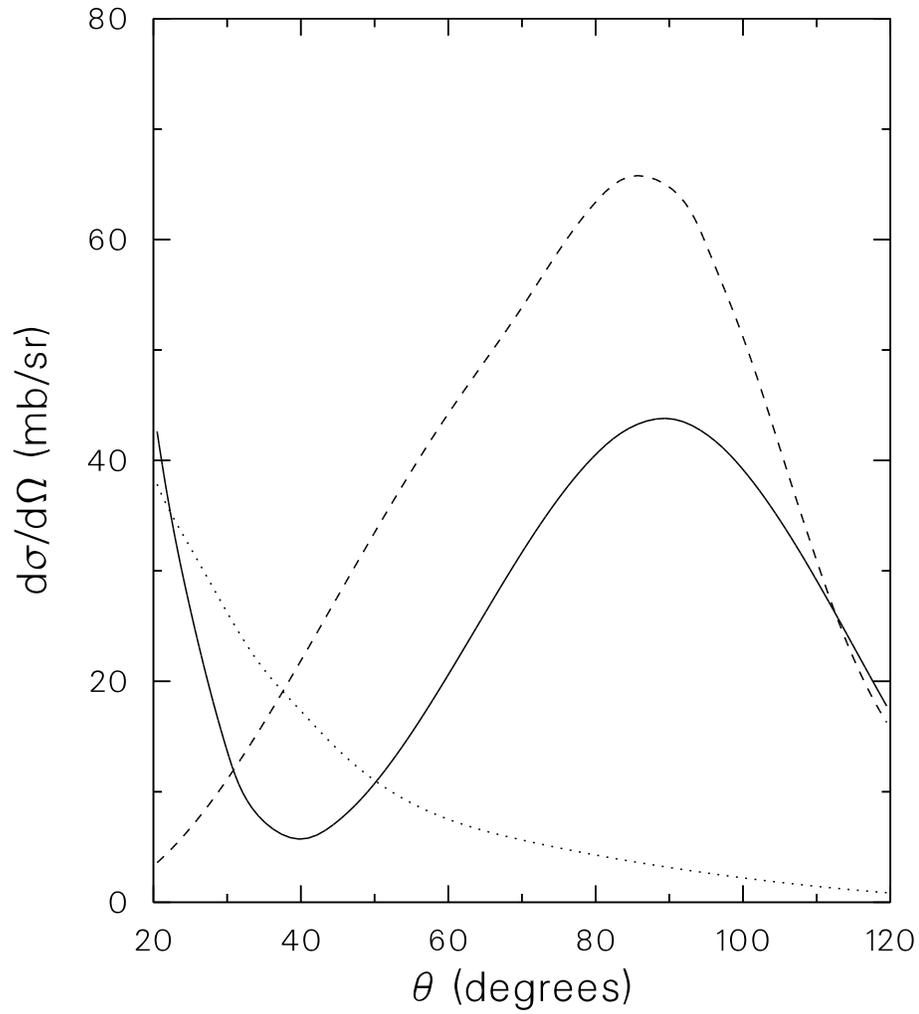
\item{Fig.~10}  
 Calculated angular distribution for the reaction 
 $^{8}$B + $^{58}$Ni at bombarding energy $E$ = 25.3 MeV, for the
 dipole transition  $p_{3/2} \to s_{1/2}$, summed over the final
 energy distribution.  The dotted line is the prediction when only
 Coulomb excitation is included, the dashed line the corresponding nuclear 
 contribution.  The solid line gives the predicted angular distribution
 including both processes. 
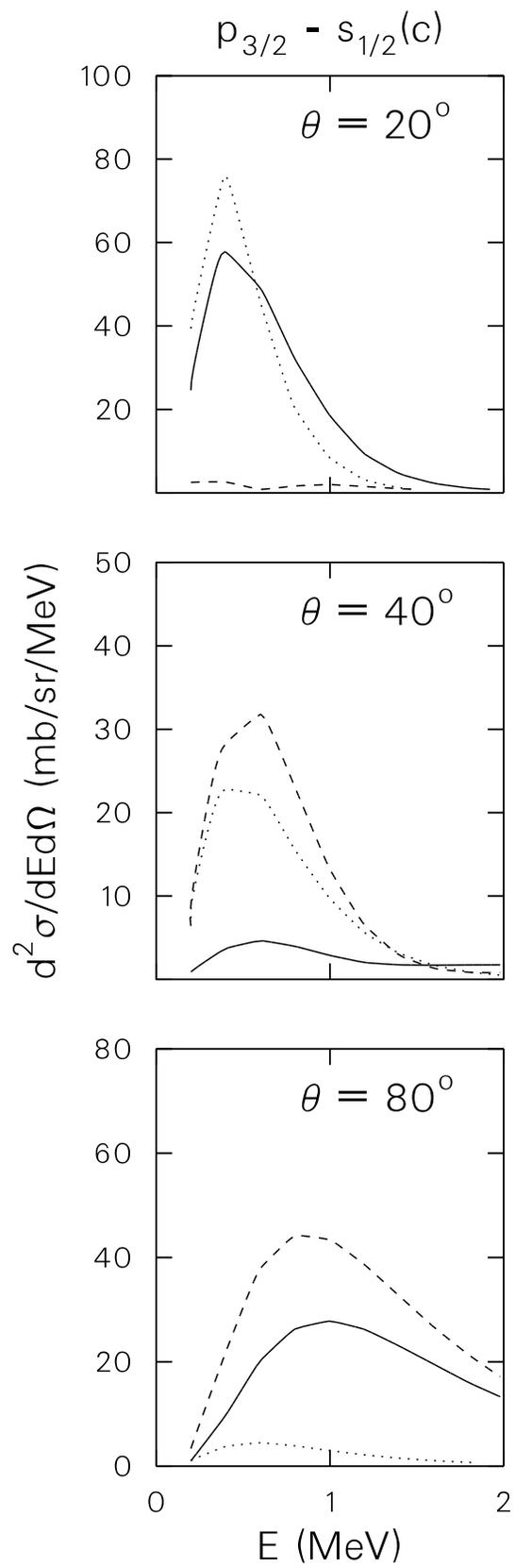
\item{Fig.~11}
 Double differential cross section for the reaction
 $^{8}$B + $^{58}$Ni at $E$=25.3 MeV as a function of the excitation energy,
 at fixed values of the scattering angle (in the c.m. system).  For
 each angle the curves refer to Coulomb excitation (dotted), nuclear
 excitation (dashed) and total (solid).
\vfill
\bye